\documentclass[reqno,11pt]{amsart}
\usepackage[foot]{amsaddr}
\usepackage{bbold}
\usepackage{charter}
\usepackage[margin=1in]{geometry}
\usepackage{graphicx}
\usepackage[colorlinks=true,linktoc=all,linkcolor=blue,citecolor=blue]{hyperref}
\usepackage{pgfplots}
\usepackage{tikz}

\theoremstyle{plain}
\newtheorem{theorem}{Theorem}[section]
\newtheorem*{theorem*}{Theorem}

\theoremstyle{definition}

\newtheorem*{definition*}{Definition}

\numberwithin{equation}{section}
\everymath{\displaystyle}
\allowdisplaybreaks

\usetikzlibrary{arrows,positioning}
\tikzstyle{b} = [rectangle, draw, node distance=1.5cm, text width=2em, text centered, rounded corners, minimum height=2em, thick]
\tikzstyle{bl} = [rectangle, draw, node distance=2.25cm, text width=2em, text centered, rounded corners, minimum height=2em, thick]
\tikzstyle{l} = [draw, -latex',thick]

\author{Robert F.~Allen\textsuperscript{1}, Cassandra Jens\textsuperscript{1}, and Theodore J. Wendt\textsuperscript{2}}
\address{\textsuperscript{1}Department of Mathematics and Statistics, University of Wisconsin-La Crosse}
\address{\textsuperscript{2}Department of Mathematics, Carroll College}

\email{rallen@@uwlax.edu, jens.cass@uwlax.edu, twendt@carroll.edu}


\title[Perturbations in Epidemiological Models]{Perturbations in Epidemiological Models\\ \textit{When Zombies Attack, We Can Survive!}}

\begin{document}

\begin{abstract}
	In this paper, we investigate the existence of stability-changing bifurcations in epidemiological models used to study the spread of zombiism through a human population.  These bifurcations show that although linear instability of disease-free equilibria may exist in a model, perturbations of model parameters may result in stability.  Thus, we show that humans can survive a zombie outbreak.
\end{abstract}

\maketitle

%%%%%%%%%%%%%%%%%%%%%%%%%%%%%%%%%%%%%%%%
\section{Introduction}

Consider a disease for which an individual, once recovered, receives ``immunity" (either death or immunity from reinfection).  To study the spread of such a disease in a healthy population, we often start by considering the Kermack-McKendrick model \cite{KermackMcKendrick:27} (commonly referred to as the SIR model).  Each individual in the population can be classified into one of three categories: 1) susceptible, 2) infectious, and 3) recovered.  Let $S(t), I(t),$ and $R(t)$ denote the number of individuals in the susceptible, infectious, and recovered categories, respectively, at time $t$. 

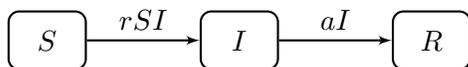
\begin{figure}[ht]
\begin{minipage}{.45\textwidth}
\begin{center}
\begin{tikzpicture}
    \node [b] (susceptibles) {$S$};
    \node [b, right=of susceptibles] (infectious) {$I$};
    \node [b, right=of infectious] (recovered) {$R$};
    \path [l] (susceptibles) -- node [above] {$rSI$}(infectious);
    \path [l] (infectious) -- node [above] {$aI$}(recovered);
\end{tikzpicture}
\end{center}
\vspace{-.1in}\caption{Kermack-McKendrick (SIR) model.}\label{Fig:SIR}
\end{minipage}%
\hspace{.4in}
\begin{minipage}{.45\textwidth}
\vspace{-4ex}
\begin{align}\label{Eq:SIR}
\begin{split}
S' &= -rSI \\
I' &= rSI - aI \\
R' &= aI
\end{split}
\end{align}
\end{minipage}%
\end{figure}

The transmission of the disease is governed by a linear mass action law, i.e., the disease is spread at a rate proportional to the number of susceptible and infectious, with individuals infected at a rate of $rSI$, where $r$ is a positive constant.  The rate of recovery is proportional to the number of infectious individuals, with individuals recovering at a rate of $aI$, where $a$ is a positive constant.  The SIR model is depicted in Figure \ref{Fig:SIR}.

This model is developed with three fundamental assumptions.  The first being the population is closed, meaning the total population $N = S(t)+I(t)+R(t)$ is constant.  Diseases to which this model is applied are short-lived in comparison to the population lifespan. The second assumption is that the individuals are homogeneously mixed, that is, every two individuals have equal probability of coming into contact.  The last assumption is that the incubation period of the disease is negligible, and so upon infection, individuals immediately move into the infectious category.  If we consider the populations to be continuous in time, we can describe the SIR model by the system of differential equations (\ref{Eq:SIR}).

\noindent The literature on the application of this model to epidemiology is too vast to cite here.  However, the interested reader is directed to \cite{BrauervandenDriesscheWu:08}, \cite{HethcoteLweisvandrnDriessche:81}, \cite{Murray:02}, and \cite{Murray:03} for further references.

More recently, the use of such mathematical models has been applied to the introduction of zombies into a human population, due in part to the rise (forgive the pun) of zombies in pop culture.  In the seminal work \cite{MunzHudeaImadSmith?:09} on modeling zombie infections, Munz et al.~used the SIR model, and its variations, to determine if humans can survive an outbreak of zombiism.   Since then, the mathematical models of zombies has taken off (see \cite{Smith?:14}).  

We will consider two models used in \cite{MunzHudeaImadSmith?:09}: the basic model, denoted by SZR, and a quarantine model, denoted by SEZQR.  Both are continuous time models with a closed total population of size $N$.  For the basic model, each individual can be classified as either susceptible, zombie, or removed zombie.  Let $S(t), Z(t)$, and $R(t)$ denote the number of susceptible, zombies, and removed zombies, respectively, at time $t$.  In this model, humans are infected, and immediately turned into zombies, at a rate of $\beta SZ$, there $\beta$ is a positive constant.  Also, zombies are removed, i.e., killed by humans, at a rate of $\alpha SZ$, where $\alpha$ is a positive constant.  In this model, the zombies are not removed through ``natural" causes such as starvation, but only through an interaction with the human population. Finally, removed zombies can be resurrected and returned to the zombie category at a rate of $\zeta R$, where $\zeta$ is a positive constant.  The model is depicted by the compartment diagram in Figure \ref{Fig:BasicModel} as well as by the system of differential equations (\ref{Eq:BasicModel}).

\begin{figure}[ht]
\begin{minipage}{.45\textwidth}
\begin{center}
\begin{tikzpicture}
    \node [b] (humans) {$S$};
    \node [b, right=of humans] (zombies) {$Z$};
    \node [b, right=of zombies] (removed) {$R$};
    \path [l] (humans) -- node [above] {$\beta SZ$}(zombies);
    \path [l] (zombies) .. controls (3.75,-.5) .. node [below] {$\alpha SZ$}(removed);
    \path [l] (removed) -- node[above] {$\zeta R$} (zombies);
\end{tikzpicture}
\vspace{-.1in}\caption{SZR model.}\label{Fig:BasicModel}
\end{center}
\end{minipage}%
\hspace{.4in}
\begin{minipage}{.45\textwidth}
\vspace{-4ex}
\begin{align}\label{Eq:BasicModel}
\begin{split}
S' &= -\beta SZ \\
Z' &= \zeta R + (\beta-\alpha) SZ \\
R' &= \alpha SZ - \zeta R
\end{split}
\end{align}
\end{minipage}%
\end{figure}

For the quarantine model, two new categories are added: an infected stage and a quarantine stage.  Susceptible individuals who are infected do not immediately turn into zombies.  They become infected but are not infectious.  Let $E(t)$ denote the number of infected (exposed) individuals at time $t$.  In this model, humans are infected at the same rate $\beta SZ$ as in the basic model.  Also, zombies are removed at the same rate $\alpha SZ$ as in the basic model.  The infected individuals become zombies at a rate proportional to the number of infected, with individuals transitioning at a rate of $\rho E$, where $\rho$ is a positive constant.  In addition, exposed individuals, as well as zombies, can be quarantined at a rate proportional to the populations.  Let $Q(t)$ denote the number of quarantined human and zombies.  The exposed humans are quarantined at a rate of $\kappa E$, where $\kappa$ is a positive constant, while zombies are quarantined at a rate of $\sigma Z$, where $\sigma$ is a positive constant.  Finally, quarantine zombies transition to the removed category at a rate of $\gamma Z$, where $\gamma$ is a positive constant.  This model is depicted by the compartment diagram in Figure \ref{Fig:LatentModel} as well as by the system of differential equations (\ref{Eq:LatentModel}).

\begin{figure}[ht]
\begin{minipage}{.45\textwidth}
\begin{center}
\begin{tikzpicture}
    \node [b] (humans) {$S$};
    \node [b, below=of humans] (infected) {$E$};
    \node [b, right=of humans] (zombies) {$Z$};
    \node [b, below=of zombies] (quarantine) {$Q$}; 
    \node [b, right=of zombies] (removed) {$R$};
    \path [l] (humans) -- node [left] {$\beta SZ$}(infected);
    \path [l] (infected) -- node [left] {$\rho E$}(zombies);
    \path [l] (zombies) .. controls (3.75,-.5) .. node [below] {$\alpha SZ$}(removed);
    \path [l] (removed) -- node[above] {$\zeta R$} (zombies);
    \path [l] (infected) -- node[below] {$\kappa E$} (quarantine);
    \path [l] (zombies) -- node[left] {$\sigma Z$} (quarantine);
    \path [l] (quarantine) .. controls (4.5,-1.5) .. node[right] {$\gamma Q$} (removed);
\end{tikzpicture}
\vspace{-.1in}\caption{SEZQR model.}\label{Fig:LatentModel}
\end{center}
\end{minipage}%
\hspace{.4in}
\begin{minipage}{.45\textwidth}
\vspace{-8ex}
\begin{align}\label{Eq:LatentModel}
\begin{split}
S' &= -\beta SZ \\
E' &= \beta SZ - (\rho+\kappa) E\\
Z' &= \rho E + \zeta R - \sigma Z - \alpha SZ \\
Q' &= \kappa E + \sigma Z - \gamma Q\\
R' &= \alpha SZ +\gamma Q - \zeta R
\end{split}
\end{align}
\end{minipage}%
\end{figure}

In both models, removed zombies are allowed to be resurrected, returning them to the zombie stage.  Thus, the immunity from reinfection present in the original SIR model has been removed.  The parameters of the two models are summarized in Table \ref{Table:Parameters}.  Estimates for the parameters are the same as those used in \cite{MunzHudeaImadSmith?:09}, and were used in the numerical solutions of the systems of differential equations.  Estimates for $\kappa, \sigma$, and $\gamma$ were not explicitly stated in \cite{MunzHudeaImadSmith?:09}, so we provide values used in the numerical analysis.

\begin{table}[ht]
\begin{center}
\begin{tabular}{c@{\hskip 2em} l@{\hskip 2em} l}
\hline
Parameter & \multicolumn{1}{c}{Description} & Estimate\\
\hline
$\beta$ & transmission rate of zombie infection & 0.0095\\
$\alpha$ & removal rate of zombies & 0.005\\
$\zeta$ & resurrection rate of zombies & 0.0001\\
$\rho$ & conversion rate of infected to zombie & 0.005\\
$\kappa$ & quarantine rate of infected humans & 0.001\\
$\sigma$ & quarantine rate of zombies & 0.001\\
$\gamma$ & removal rate of quarantined zombies & 0.0001\\
\hline
\end{tabular}
\caption{Parameters and estimates for SZR and SEZQR models.}\label{Table:Parameters}
\end{center}
\end{table}

In \cite{MunzHudeaImadSmith?:09}, the authors considered the linear stability of the disease-free equilibrium of the SZR and SEZQR models, among others, to determine the survivability of the human race.  It was shown that the disease-free equilibria are not linearly stable and verified numerically that the introduction of zombies brings about the demise of the human race.

In this paper, we wish to determine the severity of the situation.  Our strategy is to determine the sensitivity to model parameters of the linear stability of the disease-free equilibrium.  We perturb the ability to remove zombies by changing the removal mechanism to be a nonlinear mass action (as described in the next two sections).  We determine the existence of stability-changing bifurcations for the disease-free equilibria.  With this, we conclude that it is possible for humans to survive a zombie infection.  

%%%%%%%%%%%%%%%%%%%%%%%%%%%%%%%%%%%%%%%%
\section{Perturbing the Basic Model}\label{Sec:Basic}

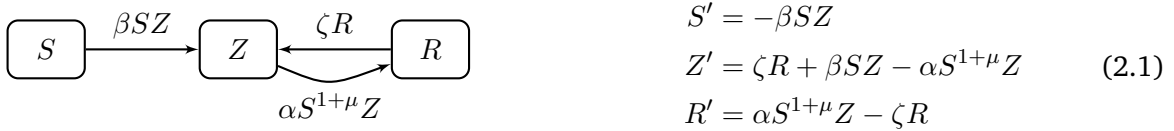
\begin{figure}[ht]
\begin{minipage}{.45\textwidth}
\begin{center}
\begin{tikzpicture}
    \node [b] (humans) {$S$};
    \node [b, right=of humans] (zombies) {$Z$};
    \node [b, right=of zombies] (removed) {$R$};
    \path [l] (humans) -- node [above] {$\beta SZ$}(zombies);
    \path [l] (zombies) .. controls (3.75,-.5) .. node [below] {$\alpha S^{1+\mu}Z$}(removed);
    \path [l] (removed) -- node[above] {$\zeta R$} (zombies);
\end{tikzpicture}
\vspace{-.1in}\caption{Peturbed SZR model.}\label{Fig:PertBasicModel}
\end{center}
\end{minipage}%
\hspace{.4in}
\begin{minipage}{.45\textwidth}
\vspace{-4ex}
\begin{align}\label{Eq:PerturbedBasicModel}
\begin{split}
S' &= -\beta SZ \\
Z' &= \zeta R + \beta SZ - \alpha S^{1+\mu}Z \\
R' &= \alpha S^{1+\mu}Z - \zeta R
\end{split}
\end{align}
\end{minipage}%
\end{figure}

In this section, we will consider a modification to the basic model as depicted in Figure \ref{Fig:BasicModel} and system (\ref{Eq:BasicModel}).  We wish to see how the stability of the disease-free equilibrium of the system is affected by perturbing the removal rate of the zombies. In this model, the ability for humans to remove zombies is being perturbed by the parameter $\mu \in (0,\infty)$, as depicted in Figure \ref{Fig:PertBasicModel} and system of ordinary differential equations (\ref{Eq:PerturbedBasicModel}).

\noindent We can reduce system (\ref{Eq:PerturbedBasicModel}) to the following system of two ordinary differential equations 
\begin{align}\label{Eq:ReducedPerturbedBasicModel}
\begin{split}
S' &= -\beta SZ \\
Z' &= \zeta (N-S-Z) + \beta SZ -\alpha S^{1+\mu}Z 
\end{split}
\end{align}
by substituting $R = N-S-Z$.

\begin{figure}[ht]
\begin{minipage}{.45\textwidth}
\begin{center}
\includegraphics[scale=.5]{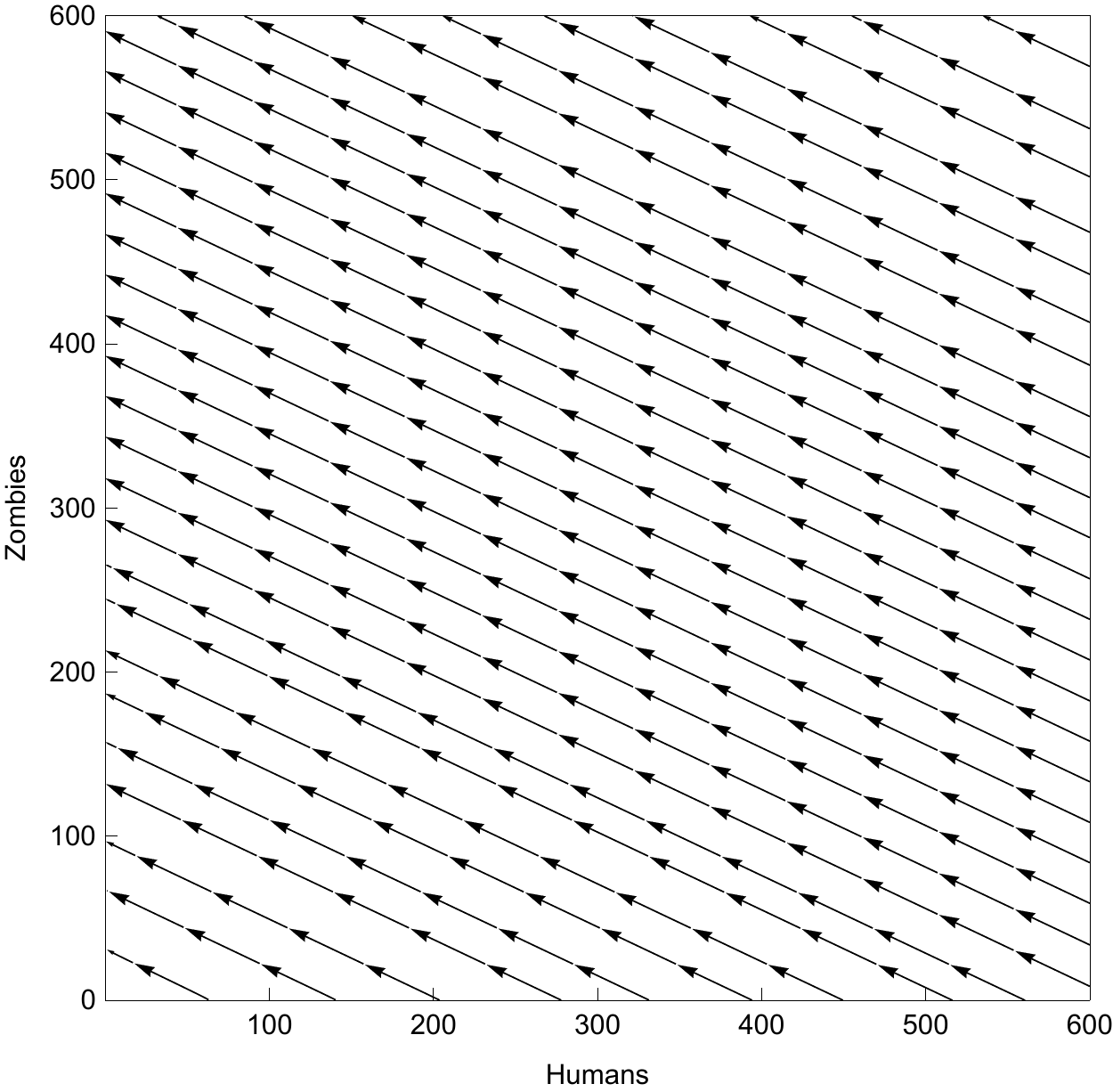}
\vspace{-.2in}\caption{Phase portrait of perturbed SZR model with $\mu=0$.} \label{Fig:PhasePortBasicModel}
\end{center}
\end{minipage}%
\hspace{.4in}
\begin{minipage}{.45\textwidth}
\begin{center}
\includegraphics[scale=.5]{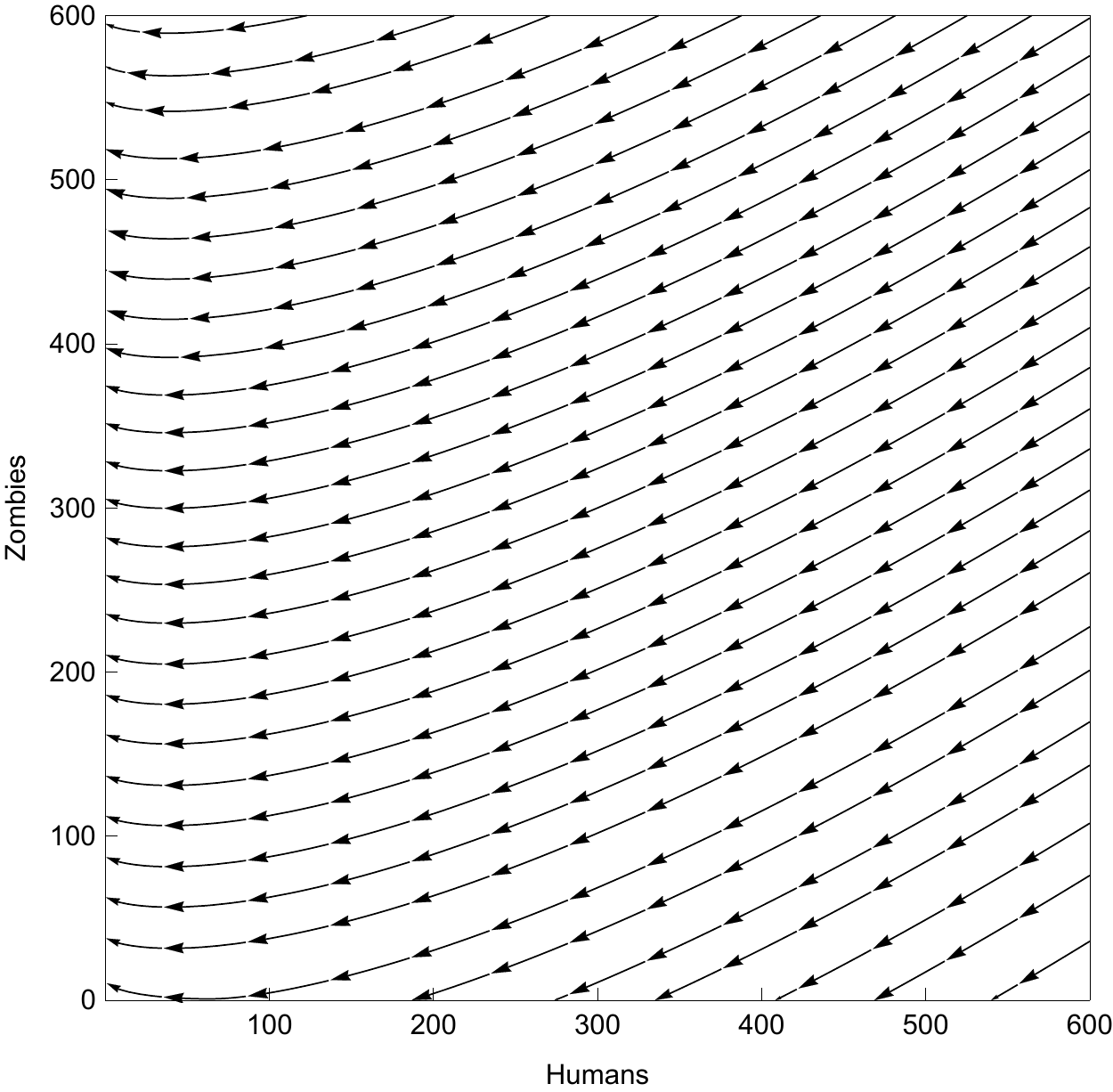}
\vspace{-.2in}\caption{Phase portrait of perturbed SZR model with $\mu=0.175$.}\label{Fig:PhasePortPertBasicModel}
\end{center}
\end{minipage}%
\end{figure}

There are two equilibria for this system, namely the epidemic equilibrium $(0,N)$ and the disease-free equilibrium $(N,0).$  If $\mu=0$, then the perturbed system is the basic model studied in \cite{MunzHudeaImadSmith?:09}.  Figure \ref{Fig:PhasePortBasicModel} shows the phase portrait of the basic model.  The trajectories in the phase portrait show that a small perturbation off the $x$-axis (the disease-free equilibrium) will result in the extinction of the human population.  

However, the phase portrait in Figure \ref{Fig:PhasePortPertBasicModel} shows that with a slight increase in $\mu$, small perturbations from the disease-free equilibrium can lead to the extinction of the zombie population.  So part of the $x$-axis (human population) is an attracting set.

We will now determine the values of $\mu$ for which the $x$-axis contains an attracting set.  We will linearize the system about the disease-free equilibrium and use the determinant-trace plane to determine linear stability.

The Jacobian of system (\ref{Eq:ReducedPerturbedBasicModel}) is given by
$$J(S,Z) = \left[\begin{matrix}
-\beta Z & -\beta S\\
-\zeta + \beta Z-\alpha(1+\mu)S^\mu Z & -\zeta + \beta S - \alpha S^{1+\mu}
\end{matrix}\right],$$ 
and evaluated at the disease-free equilibrium $(N,0)$ is 
$$J = J(N,0) = \left[\begin{matrix}
0 & -\beta N\\
-\zeta  & -\zeta + \beta N - \alpha N^{1+\mu}
\end{matrix}\right].$$
From this, we have 
\begin{align}
\det(J) &= \beta\zeta N \notag\\
\mathrm{tr}(J) &= -\zeta +\beta N - \alpha N^{1+\mu}.\notag
\end{align}

Since $\beta$ and $\zeta$ are positive, we have $\det(J) > 0$.  So for the disease-free equilibrium to be linearly stable, it must be the case that $\mathrm{tr}(J) < 0$.  This occurs when $$N^{1+\mu} > \frac{\beta N - \zeta}{\alpha}.$$  Since $\beta, \zeta,$ and $\alpha$ are all positive constants, we must impose the condition \begin{equation}\label{sys1a1}N  \geq \frac{\alpha-\zeta}{\beta}.\tag{$A1$}\end{equation}  So $\mathrm{tr}(J) < 0$ when \begin{equation}\label{Eq:basicR0}\mu > \frac{\ln\left(\frac{\beta N-\zeta}{\alpha}\right)}{\ln(N)} - 1.\end{equation}  We summarize our finding in what follows.

\begin{theorem}
The disease-free equilibrium of system (\ref{Eq:PerturbedBasicModel}) satisfying assumption (\ref{sys1a1}) is linearly stable if and only if $$\mu > \frac{\ln\left(\frac{\beta N-\zeta}{\alpha}\right)}{\ln(N)} - 1.$$
\end{theorem}

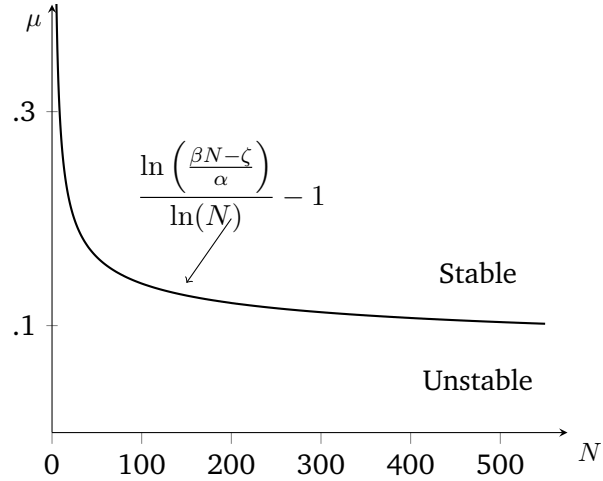
\begin{figure}[ht]
\begin{center}
\begin{tikzpicture}
    \begin{axis}[
    axis x line=bottom,
    axis y line=left,
    ylabel={\small $\mu$},
    xlabel={\small $N$},
    every axis x label/.style={at={(current axis.right of origin)},anchor=north west},
    every axis y label/.style={at={(current axis.above origin)},anchor=north east},
    axis on top,
    ytick={.1, .3, .5},
    yticklabels={.1,.3,.5},
    xtick={0, 100, 200, 300, 400, 500},
    xticklabels={0, 100, 200, 300, 400, 500},
    xtick align=outside,
    xmin=0, xmax=575, ymin=0, ymax=.4,
    samples=500,
    ]
    \addplot[smooth,thick,domain=1:550]{ln((0.0095*x-0.0001)/0.005)/ln(x) - 1};
    \draw (axis cs:200,.23) node {$\frac{\ln\left(\frac{\beta N - \zeta}{\alpha}\right)}{\ln(N)}-1$};
    \addplot[black, ->] coordinates {(200,.2) (150, .14)};
    \draw (axis cs:475,.05) node {Unstable};
    \draw (axis cs:475,.15) node {Stable};
\end{axis}
\end{tikzpicture}
\end{center}
\vspace{-.2in}\caption{Bifurcation diagram for the SZR model.}\label{Fig:BasicModelBifurcation}
\end{figure}

We will now consider the implications of this to the zombie infection.  We will analyze system (\ref{Eq:PerturbedBasicModel}) with the parameter values from Table \ref{Table:Parameters}.  First note that for $N > 1 > \frac{\alpha-\zeta}{\beta}$, and so  assumption (\ref{sys1a1}) is satisfied.  Figure \ref{Fig:BasicModelBifurcation} shows the bifurcation of instability to stability of the disease-free equilibrium.  Thus, we see that perturbations in the system parameter $\mu$ can cause the stability of the disease-free equilibrium to change from unstable to stable.  This change in the ability to remove zombies can be due to increase in skills and equipment at the disposal of the human population, for example.

%%%%%%%%%%%%%%%%%%%%%%%%%%%%%%%%%%%%%%%%
\section{Perturbing the Quarantine Model}\label{Sec:Latent}

The second model we will study is a perturbation of the SEZQR model, as described in Figure \ref{Fig:PertLatentModel} and system (\ref{Eq:PerturbedLatentModel}).

\begin{figure}[ht]
\begin{minipage}{.45\textwidth}
\begin{center}
\begin{tikzpicture}
    \node [b] (humans) {$S$};
    \node [b, below=of humans] (infected) {$E$};
    \node [b, right=of humans] (zombies) {$Z$};
    \node [b, below=of zombies] (quarantine) {$Q$}; 
    \node [b, right=of zombies] (removed) {$R$};
    \path [l] (humans) -- node [left] {$\beta SZ$}(infected);
    \path [l] (infected) -- node [left] {$\rho E$}(zombies);
    \path [l] (zombies) .. controls (3.75,-.5) .. node [below] {$\alpha S^{1+\mu}Z$}(removed);
    \path [l] (removed) -- node[above] {$\zeta R$} (zombies);
    \path [l] (infected) -- node[below] {$\kappa E$} (quarantine);
    \path [l] (zombies) -- node[left] {$\sigma Z$} (quarantine);
    \path [l] (quarantine) .. controls (4.5,-1.5) .. node[right] {$\gamma Q$} (removed);
\end{tikzpicture}
\vspace{-.1in}\caption{Perturbed SEZQR model.}\label{Fig:PertLatentModel}
\end{center}
\end{minipage}%
\hspace{.4in}
\begin{minipage}{.45\textwidth}
\vspace{-8ex}
\begin{align}\label{Eq:PerturbedLatentModel}
\begin{split}
S' &= -\beta SZ \\
E' &= \beta SZ - (\rho+\kappa) E\\
Z' &= \rho E + \zeta R - \sigma Z - \alpha S^{1+\mu}Z \\
Q' &= \kappa E + \sigma Z - \gamma Q\\
R' &= \alpha S^{1+\mu}Z +\gamma Q - \zeta R.
\end{split}
\end{align}
\end{minipage}%
\end{figure}

\noindent System (\ref{Eq:PerturbedLatentModel}) has two equilibria: the epidemic equilibrium $(0,0,N,0,0)$ and the disease-free equilibrium $(N,0,0,0,0)$.  As in the previous section, we wish to see if there are any values of $\mu$ which elicit a linearly stable disease-free equilibrium.  This would require analyzing the eigenvalues of a large matrix system.  Instead, we will use the next-generation matrix, described in \cite{DiekmannHeesterbeekRoberts:10}, to determine the basic reproduction number $R_0$ for the system.  This is interpreted as the number of secondary infections produced by an infected individual.  In the case of the SEZQR model, the number of individuals infected with zombiism by a single zombie.  The disease-free equilibrium will be linearly stable if $R_0 < 1$.  So we will use the basic reproduction number to determine if any perturbation in the SEZQR system will allow for survival of the human population.

To aid in the construction of the next-generation matrix, we will reduce the SEZQR model to a system of four ordinary differential equations replacing $R$ with $N-S-E-Z-Q$.  Thus we will consider the following system of differential equations

\begin{align}\label{Eq:ReducedPerturbedLatentModel}
\begin{split}
S' &= -\beta SZ \\
E' &= \beta SZ - (\rho+\kappa) E\\
Z' &= \rho E + \zeta (N-S-E-Z-Q) - \sigma Z - \alpha S^{1+\mu}Z \\
Q' &= \kappa E + \sigma Z - \gamma Q.\\
\end{split}
\end{align}

To construct the next-generation matrix, we focus only on the compartments which introduce new infections, or transition existing infections from compartment to compartment.  For this, we will focus on the $E', Z'$, and $Q'$ equations.  From these, we will construct the new-infection matrix $\mathcal{F}$ and the transition matrix $\mathcal{V}$ in such a way that $\left[E'\;Z'\;Q'\right]^T = \mathcal{F}-\mathcal{V}$.  For system (\ref{Eq:ReducedPerturbedLatentModel}), the new-infection and transition matrices are given by
$$ \mathcal{F} = \left[\begin{matrix}\beta SZ\\0\\0
\end{matrix}\right] \text{ and } \mathcal{V} = \left[\begin{matrix} (\rho+\kappa)E\\\alpha S^{1+\mu}Z + \sigma Z - \rho E - \zeta(N-S-E-Z-Q)\\\gamma Q-\sigma Z-\kappa E
\end{matrix}\right]. $$

Then the basic reproduction number is given by the spectral radius of $FV^{-1}$, where $F$ and $V$ are the Jacobian matrices of $\mathcal{F}$ and $\mathcal{V}$, respectively, evaluated at the disease-free equilibrium.  

Thus, we have 
$$F = \left[\begin{matrix}0 & \beta N & 0\\0 & 0 & 0\\0 & 0 & 0\end{matrix}\right] \text{ and } V = \left[\begin{matrix}\rho+\kappa & 0 & 0\\\zeta-\rho & \alpha N^{1+\mu}+\sigma+\zeta & \zeta\\ -\kappa & -\sigma & \gamma\end{matrix}\right]$$ with
$$FV^{-1} = \frac{-\beta N(\zeta\kappa+\gamma(\zeta-\rho))}{(\rho+\kappa)(\alpha\gamma N^{1+\mu} + \zeta\sigma+\gamma(\zeta+\sigma))}\left[\begin{matrix}1 & \frac{-\beta\gamma N(\rho+\kappa)}{\beta N(\zeta\kappa+\gamma(\zeta-\rho))} & \frac{\beta\zeta N(\rho+\kappa)}{\beta N(\zeta\kappa+\gamma(\zeta-\rho))}\\0 & 0 & 0\\0 & 0 & 0\end{matrix}\right].$$
Since $FV^{-1}$ is an upper-triangular matrix, we have $$R_0(\mu) = \frac{-\beta N(\zeta\kappa + \gamma(\zeta-\rho))}{(\rho+\kappa)(\alpha\gamma N^{1+\mu} +\zeta\sigma+\gamma(\zeta+\sigma))}.$$  
As we will see in equation (\ref{A1eq}), it must be the case that \begin{equation}\label{req1}\zeta\kappa + \gamma(\zeta-\rho) < 0.\tag{$A'1$}\end{equation}  Under this condition, then $R_0(\mu)$ will be positive for all values of $\mu$.  And so, to determine the values of $\mu$ for which the disease-free equilibrium is linearly stable, we determine when $R_0(\mu) < 1$.  To this end, we deduce that \begin{equation}\label{A1eq}N^{1+\mu} > \displaystyle\frac{\beta N(\zeta\kappa+\gamma(\zeta-\rho)) + (\rho+\kappa)(\zeta\sigma + \gamma(\zeta+\sigma))}{-\alpha\gamma(\rho+\kappa)}.\end{equation}  Since $\alpha\gamma(\rho+\kappa) > 0$, it must be the case that $$\beta N(\zeta\kappa+\gamma(\zeta-\rho)) + (\rho+\kappa)(\zeta\sigma+\gamma(\zeta+\sigma)) < 0,$$ which in turn requires \begin{equation}\label{req2}N > \displaystyle\frac{-(\rho+\kappa)(\zeta\sigma + \gamma(\zeta+\sigma))}{\beta(\zeta\kappa+\gamma(\zeta-\rho))}.\tag{$A'2$}\end{equation}  

Finally, for $R_0(\mu) < 1$ it must be the case that $$\mu > \frac{\ln\left(\displaystyle\frac{\beta N(\zeta\kappa+\gamma(\zeta-\rho))+(\rho+\kappa)(\zeta\sigma+\gamma(\zeta+\sigma))}{-\alpha\gamma(\rho+\kappa)}\right)}{\ln(N)} - 1,$$  which we summarize in what follows.

\begin{theorem}
The disease-free equilibrium of system (\ref{Eq:PerturbedLatentModel}), satisfying (\ref{req1}) and (\ref{req2}), is linearly stable if and only if 
$$\mu > \frac{\ln\left(\displaystyle\frac{\beta N(\zeta\kappa+\gamma(\zeta-\rho))+(\rho+\kappa)(\zeta\sigma+\gamma(\zeta+\sigma))}{-\alpha\gamma(\rho+\kappa)}\right)}{\ln(N)} - 1.$$
\end{theorem}

We will now see if this theorem can be applied to the SEZQR model, and what implications this has on the survivability of the human population.  We will analyze system (\ref{Eq:PerturbedLatentModel}) with the parameter values from Table \ref{Table:Parameters}.  First note that $\zeta\kappa+\gamma(\zeta-\rho) = -4.89\times 10^{-6} < 0$, and thus assumption (\ref{req1}) is satisfied.  Also, we have $$\displaystyle\frac{-(\rho+\kappa)(\zeta\sigma + \gamma(\zeta+\sigma))}{\beta(\zeta\kappa+\gamma(\zeta-\rho))} = 0.230546$$ So with $N > 1$, assumption (\ref{req2}) is satisfied.

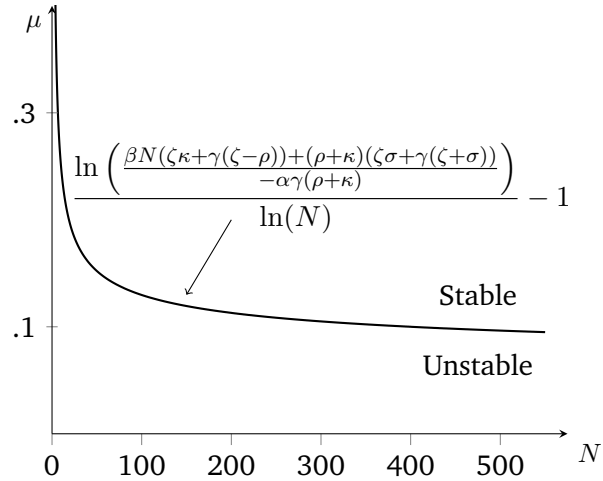
\begin{figure}[ht]
\begin{center}
\begin{tikzpicture}
    \begin{axis}[
    axis x line=bottom,
    axis y line=left,
    ylabel={\small $\mu$},
    xlabel={\small $N$},
    every axis x label/.style={at={(current axis.right of origin)},anchor=north west},
    every axis y label/.style={at={(current axis.above origin)},anchor=north east},
    axis on top,
    ytick={.1, .3, .5},
    yticklabels={.1,.3,.5},
    xtick={0, 100, 200, 300, 400, 500},
    xticklabels={0, 100, 200, 300, 400, 500},
    xtick align=outside,
    xmin=0, xmax=575, ymin=0, ymax=.4,
    samples=1000,
    ]
    \addplot[smooth,thick,domain=1:550]{ln((-.000000046455*x+.00000001071)/(-.0000000255))/ln(x)-1};
    \draw (axis cs:300,.23) node {$\frac{\ln\left(\frac{\beta N(\zeta\kappa+\gamma(\zeta-\rho))+(\rho+\kappa)(\zeta\sigma+\gamma(\zeta+\sigma))}{-\alpha\gamma(\rho+\kappa)}\right)}{\ln(N)} - 1$}; 
    \addplot[black, ->] coordinates {(200,.2) (150, .13)};
    \draw (axis cs:475,.065) node {Unstable};
    \draw (axis cs:475,.13) node {Stable};
\end{axis}
\end{tikzpicture}
\end{center}
\vspace{-.2in}\caption{Bifurcation diagram for the SEZQR model.}\label{Fig:LatentModelBifurcation}
\end{figure}

Figure \ref{Fig:LatentModelBifurcation} shows the bifurcation of instability to stability of the disease-free equilibrium.  Thus, we see that perturbations in the system parameter $\mu$ can cause the stability of the disease-free equilibrium to change from unstable to stable.  

%%%%%%%%%%%%%%%%%%%%%%%%%%%%%%%%%%%%%%%%
\section{Conclusions}\label{Sec:Conclusion}

Much study has been devoted to epidemiological models with nonlinear incidence rates, see \cite{HethcotecvandenDriessche:91}, \cite{LiMuldowney:95}, and \cite{RuanWang:03} for example.  Although the biological relevance of nonlinear incidence rates can be discussed, the introduction of the nonlinearity in this paper was not biologically motivated.  Rather, we wish to show the usefulness of perturbation analysis in the study of epidemiological models.  A linear stability analysis is often a starting point in the mathematical modeling and analysis of infectious disease.  While this analysis can offer some high-level insight into the situation, the conclusions may be sensitive to many factors.  

As was shown in the previous sections, perturbations in the removal mechanism of the zombie disease expose chaotic behavior in the system.  A linearly unstable equilibrium in the basic model can become linearly stable in this process.  The biological justification for whether such a perturbation is reasonable would be dependent on the particular disease.  In the case of zombies, the nonlinearity could be attributed to differences in the populations of humans and zombies, such as humans are organized and possibly trained while zombies are unorganized and constantly hungry.

%%%%%%%%%%%%%%%%%%%%%%%%%%%%%%%%%%%%%%%%
\section*{Acknowledgements}
The work of the second author was supported by the Wisconsin Louis Stokes Alliance for Minority Participation (National Science Foundation Grant No. 0902067) and the University of Wisconsin-La Crosse McNair Scholars Program.  
%%%%%%%%%%%%%%%%%%%%%%%%%%%%%%%%%%%%%%%%

\bibliographystyle{amsplain}
\bibliography{referenes.bib}
\end{document}